\makeatletter
\declare@file@substitution{revtex4-1.cls}{revtex4-2.cls}
\makeatother
\documentclass[twocolumn, times, tighten,twocolappendix,a4paper]{aastex63}
\usepackage{amsmath}
\usepackage{amssymb,txfonts}
\usepackage[T1]{fontenc}
\usepackage{graphicx,multirow,hyperref,url,color,xspace}
\usepackage{lineno}

\newcommand{\msun}{{\rm M}_{\sun}}

\newcommand{\g}{$\gamma$}

\newcommand{\source}{{MAXI J1820+070}\xspace}

\newbox\grsign \setbox\grsign=\hbox{$>$} \newdimen\grdimen \grdimen=\ht\grsign
\newbox\simpropbox
\setbox\simpropbox=\hbox{\raise.5ex\hbox{$\propto$}\llap
 {\lower.5ex\hbox{$\sim$}}}\ht2=\grdimen\dp2=0pt
\def\simprop{\mathrel{\copy\simpropbox}}

\begin{document}

\title{The cause of the difference in the propagation distances between compact and transient jets in black-hole X-ray binaries}
\shorttitle{Compact and transient jets}

\author[0000-0002-0333-2452]{Andrzej A. Zdziarski}
\affiliation{Nicolaus Copernicus Astronomical Center, Polish Academy of Sciences, Bartycka 18, PL-00-716 Warszawa, Poland; \href{mailto:aaz@camk.edu.pl}{aaz@camk.edu.pl}}

\author[0000-0002-8433-8652]{Sebastian Heinz}
\affiliation{Department of Astronomy, University of Wisconsin-Madison, 475 N Charter St, Madison, WI 53726, USA}

\shortauthors{A. A. Zdziarski \& Heinz}

\begin{abstract}
Accreting black-hole binaries change their properties during evolution, passing through two main luminous states, dominated by either hard or soft X-rays. In the hard state, steady compact jets emitting multiwavelength radiation are present. Those jets are usually observed in radio, and when resolved, their extent is $\lesssim\!10^{15}$ cm. Then, during hard-to-soft transitions, powerful ejecta in the form of blobs appear. They are observed up to distances of $\sim\!10^{18}$ cm, which are $\gtrsim$1000 times larger than the extent of hard-state jets. On the other hand, estimates of the accretion rates during most luminous hard states and the hard-to-soft transitions are very similar, implying that maximum achievable powers of both types of jets are similar and cannot cause the huge difference in their propagation. Instead, we explain the difference in the propagation length by postulating that the ejecta consist of electron-ion plasmas, whereas the hard-state jets consist mostly of electron-positron pairs. The inertia of the ejecta are then much higher than those of compact jets, and the former are not readily stopped by ambient media. A related result is that the accretion flow during the hard state is of Standard and Normal Evolution (SANE), while it is a Magnetically Arrested Disk (MAD) during transient ejections. The pairs in hard-state jets can be produced by collisions of photons of the hard spectrum emitted by hot accretion flows within the jet base. On the other hand, the X-ray spectra during the state transitions are relatively soft and the same process produces much fewer pairs. 
\end{abstract}

\section{Introduction}
\label{intro}

Most of accreting black-hole (BH) X-ray binaries (XRBs) are transient, having outbursts observed from radio to X-rays separated by years of quiescence. Almost all of transient BH XRBs have low-mass donors, with the mass $\lesssim\! 1\msun$, while several with either massive or undetermined donors are persistent, see \citet{Corral-Santana16}. BH XRBs show two main luminous X-ray spectral states, hard and soft, see \citet{DGK07} for a review of their X-ray properties. In radio, they show two main types of activity: compact steady jets in the hard state, and transient, discrete, jets appearing occasionally during transitions from the hard intermediate state (i.e., the softest part of the hard state, with the X-ray energy index $\alpha>1$, defined by $F_\nu\propto\nu^{-\alpha}$) to the soft state, while there is usually no or very weak radio emission during the soft state itself, see \citet{FBG04} for a review.

A striking difference between the two types of jets is in the distances to which they are observed to propagate. Compact jets appear always relatively small, in spite of the duration of their launching from the BH spanning from weeks to even years. In most cases they remain unresolved in radio. Currently, compact jets in only four BH XRBs have been resolved in radio. Three of them are MAXI J1820+070 with the deprojected jet elongation of $\ell_{\rm max} \approx\!3\times 10^{13}$\,cm at 15\,GHz \citep{Tetarenko21}, Cyg X-1 with $\ell_{\rm max} \approx\! 10^{15}$\,cm at 8\,GHz \citep{Stirling01}, and GRS 1915+105 (assuming the distance of $D\approx 8$\,kpc; \citealt{Reid14}) with $\ell_{\rm max} \approx 3\times 10^{15}({\rm 1\,GHz}/\nu)$ \citep{Dhawan00}. Then, MAXI J1836--194 has an only rough (due to the distance and inclination of this source being uncertain) estimate of the length of $\ell_{\rm max} \sim\! 10^{16}$\,cm at 2.3\,GHz \citep{Russell15}, which was obtained in only one out of several VLBA observations after subtracting the point source. Theoretically, we expect $\ell_{\rm max} \simprop \nu^{-1}$ \citep{BK79}. Then, not a single discrete extension of a hard-state jet was detected moving at large distances from the core. 

On the other hand, transient jets are observed as discrete moving ejecta at much larger distances, up to $\sim\! 10^{17}$--$10^{18}$\,cm, in spite of the duration of their launching being of the order of a day (e.g., \citealt{Carotenuto21, Carotenuto24}). Often both the approaching and the receding components are seen (e.g., \citealt{MR94, Fender99, Bright20}), and they are sometimes detected even at distances $>\! 10^{18}$\,cm \citep{Carotenuto21, Carotenuto24}. Both types of jets often appear during the same observational campaigns, e.g., for \source and MAXI J1348--630. The cause of the difference in the propagation distances has remained unexplained. So far, this type of activity has been detected only from transient sources (i.e., showing outbursts separated by years of quiescence), and not from persistent sources.

We need to mention another small class of BH XRBs that show parsec-scale persistent radio jets. We know two such sources, 1E 1740.7--2942 \citep{Mirabel92, Luque15} and GRS 1758--258 \citep{Rodriguez92, Marti02, Marti17}. Both are persistent X-ray sources. Then, the jet in the persistent source Cyg X-1 may power an interstellar shell distant by $\sim$10\,pc \citep{Gallo05}, though it remains uncertain \citep{Sell15}, and the hypothetical large-scale jet connecting to the shell is invisible. Then, there is yet another case of the persistent source, namely Cyg X-3, which has major radio outbursts in its soft state (e.g., \citealt{Koljonen10}). Here, we will not consider those cases, and we will concentrate on the differences between compact and transient jets in transient BH XRBs only.

Another difference between the two types of jets is in their radio spectra. The radio emission of the compact jets is flat or inverted \citep{Fender01b}, with $\alpha\lesssim 0$, which is well explained by the partially synchrotron self-absorbed jet model of \citet{BK79}, see, e.g., \citet{Tetarenko21}, \citet{Zdziarski22a}, while that of the transient jets is steep, $\alpha\sim 0.5$ (e.g., \citealt{Rodriguez95, Carotenuto21}), characteristic of optically thin synchrotron emission by power-law electrons. The emitting regions of compact jets are continuously replenished and stay optically thick up to some wavelength-dependent distances, while transient ejecta rapidly expand to large lateral sizes (e.g., \citealt{MR94, Rushton17}), causing the self-absorption optical depth to be low.

On the other hand, the bulk Lorentz factors of both types of jets appear relatively similar. These factors for compact jets remain difficult to constrain, with only rough estimates of $\Gamma\gtrsim 1.5$--2 \citep{Stirling01, Casella10}. \citet{Tetarenko19} gave $\Gamma= 2.6^{+0.8}_{-0.6}$ for Cyg X-1 (note the caveats given in \citealt{Z_Egron22}). In \source, \citet{Zdziarski22a} found $\Gamma\approx 1.5$--4. For transient jets, some jets are observed to decelerate, and we list here the estimates for the initial Lorentz factor, $\Gamma_0$. \citet{Corbel05} and \citet{Steiner12b} gave $\Gamma_0\approx 1.4$--1.6 in H1743--322, \citet{Steiner12} found $\Gamma_0\gtrsim 1.6$ with no upper limit for XTE J1550--564, \citet{Wang03} obtained a good fit for the same object with $\Gamma_0=3$, \citet{Wood21} found $\Gamma_0>2.1$ and $\Gamma_0\approx 1.05\pm 0.02$ for two ejecta of \source, and \citet{Zdziarski23a} obtained $\Gamma_0\approx 1.8\pm 0.1$ in MAXI J1348--630. Thus, the transient jets appear to have the Lorentz factors similar to those of the compact steady jets.

Also, existing estimates of the powers, $P_{\rm j}$, of the compact and transient jets are relatively similar. A theoretical upper limit on $P_{\rm j}$ is $\sim \dot M_{\rm accr} c^2$, where $\dot M_{\rm accr}$ is the accretion rate (\citealt{Davis20} and references therein). The accretion luminosities, $L$, of the hard state before the state transition and during it are quite similar, and the accretion efficiency, $\epsilon$, is theoretically predicted to increase from the hard state to the soft one \citep{YN14}. Then $\dot M_{\rm accr}=L/(\epsilon c^2)$ is unlikely to substantially increase at the hard-to-soft transition. For compact jets, published estimates are mostly in the $P_{\rm j}\sim 10^{37-39}$\,erg/s range, e.g., \citet{Zdziarski22a}. In the case of transient jets, there were some estimates of extreme powers, $P_{\rm j}\sim 10^{41}$\,erg/s, in particular those of \citet{MR94} and \citet{Carotenuto22}. However, these values were revised and found much lower by \citet{Zdziarski14d} and \citet{Zdziarski23a}. In the recent study of \citet{Carotenuto24}, the kinetic energy of three transient jets were calculated, and they also imply a $P_{\rm j}\sim 10^{37-39}$\,erg/s range.

In a recent work, \citet{SZ23} proposed that the jets produced during the hard-to-soft state transitions are transient because of an increase of the accretion rate, $\dot M_{\rm accr}$, associated with the transitions. If the jet before the transition was launched from a magnetically-arrested accretion disk (MAD; \citealt{McKinney12}) and the accumulated magnetic flux remains constant during the transition (see fig.\ 1 in \citealt{SZ23}), the system can cease to be in the MAD state above certain $\dot M_{\rm accr}$, which, in turn, can decollimate the outflow and stop the jet production. The discrete ejection would correspond then to a part of the compact jet produced just before the decollimation. This picture implies that the transient jets are similar to the compact ones except for a shorter launching time. This, however, does not explain the dramatic difference in the propagation distances between the compact and transient jets, which thus is still unknown. In this work, we attempt to explain this difference. We concentrate our study on the case of \source, whose detailed observations allow relatively reliable estimates of the jet parameters for both the compact jet and the transient ones.

\section{The case of \source}
\label{maxi}

\subsection{Measurements and estimates}
\label{estimates}

The deprojected elongation of the jet resolved at 15 GHz by VLBA on MJD 58193 (5 days after the source discovery in X-rays, \citealt{Kawamuro18}) for the distance of $D\approx 3.0\pm 0.3$\,kpc (as measured by \citealt{Atri20}) and the inclination of $64\degr$ \citep{Wood21} is $\ell_{\rm max}\approx (2.6\pm 0.1)(D/3\,{\rm kpc}) 10^{13}$\,cm \citep{Tetarenko21}. Extrapolating it to MJD 58220 based on the radio flux (with a model based on \citealt{BK79}), the estimated length for that day was $\ell_{\rm max} \approx 4\times 10^{13}$\,cm \citep{Zdziarski22a}. This also approximately agrees with the time-lag estimates of \citet{Zdziarski22a}, see their fig.\ 7, as well as with the estimates based on the breaks in the power spectra \citep{Tetarenko21, Zdziarski22a}. Then, the size estimates based on the method of core shift are still lower by a factor of several \citep{Prabu23}. 

On the other hand, the main approaching ejection in that source, launched during the hard-to-soft state transition (around MJD 58306; \citealt{Wood21}) was detected in X-rays at the distances from the compact object as large as $\approx\! 6\times 10^{17}$\,cm \citep{Espinasse20}, which is $>$4 orders of magnitude longer than the measured length of the hard-state jet. This is in spite of the similar jet Lorentz factors, see below, and in spite both types of jets apparently launched in the same directions. 

The Lorentz factor of the compact jet was estimated as $\Gamma\approx$1.5--4 \citep{Zdziarski22a}. Then, \citet{Carotenuto24} estimated the initial Lorentz factor as $\Gamma_0\approx 2.6^{+0.5}_{-0.4}$ for the main transient ejection. Thus, there is no indication on which of the jets has the higher $\Gamma$. 

The main method for estimating the jet power in steady, compact, jets is to use the observed synchrotron flux to calculate the flux of the emitting electrons and positrons (e$^\pm$). By assuming the ratio of the energy densities in the e$^\pm$ to that in the magnetic field (the equipartition parameter), we can calculate the power in the e$^\pm$ and the field, $P_{B{\rm e}}$. Another component is the power in ions, $P_{\rm i}$, which depends on the fraction of e$^\pm$ pairs and on the number of non-emitting electrons at low energies. In the case of \source at the absence of non-emitting electrons, 
\begin{align}
&P_{B{\rm e}}\sim 10^{37}\,\frac{\rm erg}{\rm s},\quad
P_{\rm i}\sim 10^{39}\left(1-\frac{2 n_+}{n_+ +n_-}\right) 
\frac{\rm erg}{\rm s},\nonumber\\ 
&P_{\rm j,compact}=P_{\rm Be}+P_{\rm i}
\label{Pjetc}
\end{align}
(see eqs.\ 45--46 of \citealt{Zdziarski22a}), and $P_{\rm i}\approx 0$ for a pure pair twin jet. Those calculations were based on the fast-timing and spectral results of \citet{Tetarenko21}, and the underlying observations were done during the luminous hard state of that source (on MJD 58220, 2018-04-12). 

Observational estimates of the jet power can be compared to theoretical predictions for the jet power. The most efficient such model known as yet is that based on the extraction of the rotation power of the BH \citep{BZ77}. In this model, $P_{\rm j}\approx k a_*^2 \Phi_{\rm BH}^2$, where $k$ is a constant \citep{Tchekhovskoy15}, $a_*$ is the BH spin parameter, and $\Phi_{\rm BH}$ is the magnetic flux threading the BH. Then, the maximum possible $\Phi_{\rm BH}\equiv \Phi_{\rm MAD}$ corresponds to the balance between the magnetic pressure and the ram pressure of the accreting matter, i.e., a Magnetically Arrested Disk (MAD; \citealt{BK74, Narayan03}). Based on numerical simulations of jets \citep{Tchekhovskoy11, Davis20}, it has been found to correspond to (for both the jet and counterjet) 
\begin{equation}
P_{\rm MAD}\approx 1.3 a_*^2 \dot M_{\rm accr} c^2,
\label{Pmax}
\end{equation}
which gives the upper limit on the possible jet power, $P_{\rm j}\leq P_{\rm MAD}$. Assuming isotropy, we estimate 
\begin{equation}
\dot M_{\rm accr} c^2 = 4 \pi F_{\rm accr} D^2/\epsilon, 
\label{Mdotc2}
\end{equation}
where $F_{\rm accr}$ is the bolometric accretion flux, and $\epsilon$ is the accretion efficiency. 

\citet{Shidatsu19} found the unabsorbed 1--100\,keV flux in the luminous hard state to be between $\approx$0.5 and $\approx\! 1.4\times 10^{-7}$\,erg\,cm$^{-2}$\,s$^{-1}$. The 0.1--200\,keV flux as observed by XMM and INTEGRAL in the luminous hard state on MJD 58220 (for which the hard-state jet power was estimated above) was $\approx\! 1.1\times 10^{-7}$\,erg\,cm$^{-2}$\,s$^{-1}$ (J. Rodi, private communication; \citealt{Rodi21}). We find the estimated bolometric flux of $\approx\! 1.2\times 10^{-7}$\,erg\,cm$^{-2}$\,s$^{-1}$, corresponding to $L_{\rm bol}\approx 1.3(D/3\,{\rm kpc})^2 10^{38}$\,erg\,s$^{-1}$. For the BH mass of $M=6.8\msun$ \citep{Torres20, Mikolajewska22} and for the H mass fraction of $X=0.7$, this $L_{\rm bol}$ represents 13\% of the Eddington luminosity. It implies $\dot M_{\rm accr} c^2 \approx 1.3\times 10^{39}(D/ 3\,{\rm kpc})^2 (\epsilon/ 0.1)^{-1}$\,erg\,s$^{-1}$. The initial estimates of $a_*$ using the continuum method were $a_*\approx 0.2^{+0.2}_{-0.3}$ \citep{Guan21} and $a_*\approx 0.14\pm 0.09$ \citep{Zhao21}, or even $a_*<0$ \citep{Fabian20} at the binary inclination of $\approx$66--$81\degr$ \citep{Torres20}. On the other hand, \citet{Bhargava21} obtained $a_*\approx 0.8$ from timing using the relativistic precession model. Similarly, \citet{Banerjee24} obtained $a_*=0.77\pm 0.21$. Scaling the power to $a_*=0.8$ gives 
\begin{equation}
P_{\rm MAD,compact}\approx 1.0(a_*/0.8)^2(\epsilon/0.1)^{-1} 10^{39}\,{\rm erg\,s}^{-1}.
\label{PMADc}
\end{equation}
Thus, the estimate for the jet compact power based on the synchrotron emission of Equation (\ref{Pjetc}) in the absence of pairs is approximately equal to $P_{\rm MAD,compact}$ at $a_*\approx 0.8$. On the other hand, $P_{\rm j,compact}\ll P_{\rm MAD,compact}$ for the same estimate if the jet composition is dominated by pairs.

\setlength{\tabcolsep}{4pt}
\begin{table}\centering
\caption{Comparison of the source properties during the compact and transient jet launching in \source} 
\vskip -0.4cm                               
\begin{tabular}{ccccc}
\hline
Jet Type&  $L_{\rm bol}$ & $\Gamma$ & $P_{\rm MAD}$ & $\ell_{\rm max}$\\
 & erg\,s$^{-1}$ && erg\,s$^{-1}$ &cm\\ 
\hline
Compact & $1.3\times 10^{38}$ & 1.5--4 & $1.0 \left(\frac {a_*}{0.8}\right)^2 \frac{0.1} {\epsilon} 10^{39}$ &$4\times 10^{13}$\\
Transient & $1.8\times 10^{38}$ &   $2.6^{+0.5}_{-0.4}$ & $1.5\left(\frac {a_*}{0.8}\right)^2 \frac {0.1}{\epsilon} 10^{39}$ & $6\times 10^{17}$\\
\hline
\end{tabular}
\label{comparison}
\tablecomments{$D=3$\,kpc is assumed, and $\ell_{\rm max}$ for the compact jet is based on the radio imaging at 15\,GHz. We see that while both the accretion and jet properties appear similar, the maximum distances travelled are very different.} 
\end{table}

Just before and during the main transient ejection from \source (estimated to occur on MJD 58306, 2018-07-07; \citealt{Wood21}), the source was in the intermediate state \citep{Shidatsu19}, which spectrum approximately consists of a disk blackbody followed by a steep power law with $\alpha>1$. The 0.1--78\,keV flux on MJD 58306 was measured as $\approx\! 1.62\times 10^{-7}$\,erg\,cm$^{-2}$\,s$^{-1}$ \citep{Fabian20}, which, given the shape of the spectrum is only slightly lower than the estimated bolometric flux, $\approx 1.7\times 10^{-7}$\,erg\,cm$^{-2}$\,s$^{-1}$. The bolometric luminosity is then $L_{\rm bol}\approx 1.8(D/3\,{\rm kpc})^2 10^{38}$\,erg\,s$^{-1}$ (which is 18\% of the Eddington luminosity), and 
\begin{equation}
P_{\rm MAD,transient}\approx 1.5(a_*/0.8)^2(\epsilon/0.1)^{-1} 10^{39}\,{\rm erg\,s}^{-1},
\label{PMADt}
\end{equation}
similar to $P_{\rm MAD,compact}$, Equation (\ref{PMADc}). The intermediate state has a stronger disk component than the hard state, indicating the disc inner radius, $R_{\rm in}$, is closer to the innermost stable circular orbit than in the hard state. Since the efficiency of thin disks is $\approx R_{\rm g}/(2 R_{\rm in})$ \citep{SS73} and the efficiency of hot flows is generally lower than that \citep{YN14}, $\epsilon$ is expected to be higher in the intermediate state than in the hard state. Then, the estimated $\dot M_{\rm accr} c^2$ and $P_{\rm MAD,transient}$ can be even more similar to that for the hard state. Table \ref{comparison} compares the accretion luminosity, the jet properties and the maximum distance travelled for the two types of jet launching. 

\subsection{Implications}
\label{implications}

The energy content of transient jets can be estimated from modelling their motion in the surrounding medium (e.g., \citealt{Wang03, Steiner12, Steiner12b, Zdziarski23a, Carotenuto24}), which can be converted to the jet power if the duration of the ejection event can be estimated. However, those estimates scale with the unknown density of the surrounding medium, which appears much lower than the density of a warm ISM, $n\ll 1$\,cm$^{-3}$, if the jet power is limited from above by $\dot M_{\rm accr} c^2$ \citep{Heinz02}. Since the densities of the surrounding media remain unknown, we can only place upper limits on the jet power, equal to that of Equation (\ref{Pmax}).

Taking $P_{\rm j,transient}=P_{\rm MAD,transient}\approx 1.5\times 10^{39}$\,erg\,s$^{-1}$, Equation (\ref{PMADt}), the kinetic energy of the one-sided ejection of \source estimated using the ejection duration of 7 h (as in \citealt{Carotenuto24}) becomes $E_0\approx 2\times 10^{43}$\,erg, which, given the Lorentz factor estimated by \citet{Carotenuto24}, gives the single mass of $E_0/[(\Gamma_0-1)c^2]\approx 1.4\times 10^{22}$\,g. If the ejection mass were dominated by low-energy e$^\pm$, there would $1.4\times 10^{49}$ pairs in both ejecta. For the 7 h ejection duration, this requires the total pair production rate of $N_+\approx 6\times 10^{44}$\,s$^{-1}$. This is more than four orders of magnitude larger than $\dot N_+\sim 2\times 10^{40}$\,s$^{-1}$ estimated for the hard state of this source by considering pair production by accretion photons ($\gamma\gamma \rightarrow {\rm e}^+{\rm e}^-$) within the base of the compact jet \citep{Zdziarski22a}. Furthermore, the X-ray spectrum of that intermediate state \citep{Fabian20} was much softer than that of the hard state, and the number of photons available for pair production was correspondingly much lower. Thus, we conclude that the ejection composition was dominated by ions, similar to the case of MAXI J1348--630 \citep{Zdziarski23a}. 

On the other hand, if the composition of compact jets were dominated by ions, its power would be at the MAD limit, compare Equations (\ref{Pjetc}) and (\ref{PMADc}), which would in turn leave the huge difference between the propagation lengths of the compact and transient jets unexplained. However, \citet{Zdziarski22a} showed that the rate of pair production in the hard state (see the paragraph above) approximately equals the rate of the flow of the synchrotron-emitting leptons. Thus, the jet in that state can easily be dominated by relativistic e$^\pm$ pairs, with only few ions. The same conclusion was reached for Cyg X-1 \citep{Z_Egron22} and the radio galaxy 3C 120 \citep{Zdziarski22c}. Then, the hard-state jet power based on the synchrotron emission and assuming pair dominance is $P_{\rm j,compact}\ll P_{\rm MAD,compact}$. This, in turn, implies that the hard-state accretion flow contains a magnetic flux much lower than that of the MAD, $\Phi\ll \Phi_{\rm MAD}$, and thus it is of the Standard and Normal Evolution (SANE) type \citep{Narayan12b}. The differences in both the jet composition and the power can then explain the difference in the jet propagation length, see Section \ref{propagation} below. This conclusion also agrees with that of \citet{Fragile23}, based on theoretical modelling of quasi-periodic oscillations, that the luminous hard state cannot be MAD.  

Summarizing the above, we find that the jet Lorentz factors, $\dot M_{\rm accr} c^2$, and the maximum possible jet powers, $P_{\rm MAD}$, are similar for both kinds of jets, see Table \ref{comparison}. Furthermore, if the compact jets were dominated by ions, their powers would be similar to $P_{\rm MAD,compact}$, which is unlikely. The power of transient jets cannot be uniquely determined, but given their large propagation distances, it is much higher than that of the compact jets, and likely at the MAD limit. 

Thus, considering possible differences between the two kinds of jets that would explain the difference in their propagation, we postulate they lie in both the composition and the jet power. Transient jets are clearly dominated by normal plasma with at most few pairs, and their power appears to be at the MAD limit.  In contrast, compact jets are very likely dominated by pairs, though this is not proven by their modelling. Their power is then low and much below the MAD limit since it is dominated by the component due to relativistic pairs and magnetic field (with an at most small contribution from ions, cf.\ Equation \ref{Pjetc}). Consequently, the magnetic flux threading the BH in the hard state has to be well below the MAD limit. This, in turn, can explain the very limited distances they propagate, see Section \ref{propagation}.

The power of the transient jet is, as discussed above, uncertain since it scales with the unknown density of the surrounding medium. However, given their very long propagation distances, it is likely that their power is at the MAD limit, with $P_{\rm MAD,transient}\approx 1.5(a_*/0.8)^2 10^{39}$\,erg\,s$^{-1}$ in the case of \source. For $a_*=0.8$, is two orders of magnitude larger than the $P_{\rm j,compact}$ for the jet dominated by pairs.

We also mention a possible mechanism \citep{Thomas22} of the ejections of transient jets, which happen on the time scale of a day or less, much shorter than the duration of the launching of compact jets. An accumulating magnetic field flux can stop the accretion at some radius. Then, matter accumulates outside this radius, and finally breaks through it, causing the advected magnetic field to accumulate on the BH, reaching the MAD state, during which the infalling matter gets ejected.

Finally, we consider correspondence to jets in radio loud AGNs. Those jets have two main types, FR I and FR II \citep{FR74}. FR I sources show jets whose luminosity decreases as the distance from the central BH and have lower radio luminosity than FR II jets, suggesting they are counterparts of the compact jets in XRBs. While their total spectra are steep, their core spectra are flat on average, with $\langle\alpha\rangle\approx 0$ \citep{Yuan18}. Then, FR II jets are edge-brightened, with luminous radio lobes, and could be associated with the transient jets in XRBs. However, their radio spectra are typically flat, unlike those of the transient ejecta. Furthermore, \citet{Sikora20} have shown that the FR II jet composition appears to be dominated by e$^\pm$ pairs, again unlike the case of the transient ejecta. Also, both types of AGN jets are long-lasting, different from the transient XRB ejections. Thus, there seems to be no simple correspondence between the two FR types in AGNs and the two types of jets in XRBs.

\section{Propagation of jets}
\label{propagation}

The propagation of transient jets appears to be well described by the widely-used formalism of \citet{Wang03}, which then requires the density of the surrounding medium to be very low. Within that formalism, we have derived the density of (similar to, but more accurate than eq.\ 5 of \citealt{Heinz02}) 
\begin{equation}
n\approx \frac{3 E_0 (1-k) \Gamma_0}{\pi  m_{\rm p}c^2 l_k^3 s (\Gamma_0 -1) \phi^2 (k^2 \Gamma_0^2-1)},
\label{n}
\end{equation}
where $l_k$ is the distance at which the Lorentz factor of the ejection is reduced to $k\Gamma_0$ ($1/\Gamma_0<k<1$), $\phi$ is the half-opening angle, $s\approx 0.7$ \citep{Wang03}, and $m_{\rm p}$ is the proton mass. In the derivation, we made a simplifying assumption that the Lorentz factor of the shock front equals that of the ejection. The stopping length is then
\begin{equation}
l_k\approx
\left[\frac{3 E_0 (1-k) \Gamma_0}{\pi n m_{\rm p}c^2 s (\Gamma_0 -1) \phi^2 (k^2 \Gamma_0^2-1)}\right]^{1/3}.
\label{loading}
\end{equation}
For $E_0=2\times 10^{43}$\,erg\,s$^{-1}$ (see Section \ref{maxi}), $k=0.5$, $\Gamma_0=2.6$ and $n\lesssim 10^{-3}$\,cm$^{-3}$ \citep{Carotenuto24}, we obtain $l_{1/2}\gtrsim 4\times 10^{17}$\,cm, which agrees well with the observations. 

In the case of compact jets dominated by e$^\pm$, the pairs in the jet may first lose most of their energy, strongly reducing the mass. This may provide the required reduction of the kinetic energy. Indeed, \citet{Zdziarski22a} found that the synchrotron power of \source is very similar to $P_{B{\rm e}}$. If $P_{\rm j}\approx P_{B{\rm e}}$ (i.e., the bulk kinetic power of ions can be neglected), the e$^\pm$ will indeed lose most of their internal energy during the propagation. We will then have a dark jet, possibly propagating to large distances. In the presence of both synchrotron and adiabatic losses, the relativistic electrons lose all their energy. In the absence of adiabatic losses, they reach a terminal Lorentz factor $>1$ \citep{Kaiser06}.

Then (for the case of adiabatic losses) the jet power will be equal to that of the energy flow of the cold pairs and the magnetic field. For \source, it can be estimated as $\approx 4\times 10^{34}$\,erg\,s$^{-1}$ \citep{Zdziarski22a} provided the magnetic field is effectively dissipated, maintaining the equipartition. In order to find the jet kinetic energy, we need to estimate the characteristic time this power operates before the jet is slowed down. The maximum estimate of this time is $l_k/c$, $k\sim 1/2$. Inserting $E_0=(1/2)P_{\rm j} l_k/c$ into Equation (\ref{loading}) gives 
\begin{equation}
l_k\approx
\left[\frac{3 P_{\rm j} (1-k) \Gamma_0}{2\pi n m_{\rm p}c^3 s (\Gamma_0 -1) \phi^2 (k^2 \Gamma_0^2-1)}\right]^{1/2}.
\label{stopping}
\end{equation}
For $P_{\rm j}=4\times 10^{34}$\,erg\,s$^{-1}$, $k=0.5$, $\Gamma_0=2.6$ and $n=10^{-3}$\,cm$^{-3}$, we still find a large $l_{1/2}\approx 5\times 10^{16}$\,cm. Thus, the jet made of cold pairs would still be stopped at a large distance from the BH, and we need some additional processes to efficiently stop it. However, the main uncertainty in this estimate is in the assumption of the time the jet kinetic energy effectively accumulates of $E_0=(1/2)P_{\rm j} l_k/c$. If the effective $E_0$ is much lower, the jet will stop at a much closer distance. An additional effect increasing the efficiency of the jet stopping is its lateral expansion, increasing $\phi$. The jet stopping process will be considered in detail in Heinz et al.\ (in preparation). 
 
\section{Conclusions}
\label{conclusions}

We have confirmed that the accretion rate during launching of compact jets during the luminous hard state is very similar to that in the case of transient jet launching, which in turn, happens during transitions from the hard state to the soft one. This implies that the maximum possible jet powers, achieved during the MAD, are very similar. We used a specific example of the very well-studied BH XRB, \source. The estimated/measured bulk Lorentz factors of both the compact and transient jets also appear similar. Furthermore, the compact jet power estimated from its synchrotron emission assuming the absence of e$^\pm$ pairs would be at the MAD limit. 

This would then leave unexplained the striking difference between the two types of jets, with transient jets propagating to much larger distances than the compact ones. We postulate here that compact jets consist mostly of e$^\pm$ pairs, which strongly reduces both their powers and inertia estimated from their synchrotron emission. A sufficient number of pairs can be produced by photon-photon collisions of hard X-rays/soft \g-rays emitted by the accretion flow (as found in earlier works). Then, the lower jet power together with a high accretion rate imply that the hard-state luminous accretion flow has the magnetic flux much below the MAD limit, and it is consequently of the SANE type. This confirms the conclusion of \citet{Fragile23} that the hard state is not MAD. 

Compact jets then propagate and lose their internal energy, further reducing their power. This can then explain the compact jets propagating to much shorter distances than the transient jets. Propagation of compact jets in BH XRBs will be studied in detail in Heinz et al. (in preparation). 

\section*{Acknowledgements}
We acknowledge the referee for valuable comments. We thank J.-P. Lasota, J. Miller-Jones and T. Russell and M. Sikora for discussions and J. Rodi for providing us with the broad-band X-ray luminosity of \source as observed by XMM and INTEGRAL. We acknowledge support from the Polish National Science Center grants 2019/35/B/ST9/03944 and 2023/48/Q/ST9/00138, and from the Copernicus Academy grant CBMK/01/24. 

\bibliographystyle{aasjournal}
\bibliography{../allbib} 

\begin{thebibliography}{}
\expandafter\ifx\csname natexlab\endcsname\relax\def\natexlab#1{#1}\fi
\providecommand{\url}[1]{\href{#1}{#1}}
\providecommand{\dodoi}[1]{doi:~\href{http://doi.org/#1}{\nolinkurl{#1}}}
\providecommand{\doeprint}[1]{\href{http://ascl.net/#1}{\nolinkurl{http://ascl.net/#1}}}
\providecommand{\doarXiv}[1]{\href{https://arxiv.org/abs/#1}{\nolinkurl{https://arxiv.org/abs/#1}}}

\bibitem[{{Atri} {et~al.}(2020){Atri}, {Miller-Jones}, {Bahramian}, {Plotkin},
  {Deller}, {Jonker}, {Maccarone}, {Sivakoff}, {Soria}, {Altamirano},
  {Belloni}, {Fender}, {Koerding}, {Maitra}, {Markoff}, {Migliari}, {Russell},
  {Russell}, {Sarazin}, {Tetarenko}, \& {Tudose}}]{Atri20}
{Atri}, P., {Miller-Jones}, J.~C.~A., {Bahramian}, A., {et~al.} 2020, \mnras,
  493, L81, \dodoi{10.1093/mnrasl/slaa010}

\bibitem[{{Banerjee} {et~al.}(2024){Banerjee}, {Dewangan}, {Knigge},
  {Georganti}, {Gandhi}, {Mithun}, {Saikia}, {Bhattacharya}, {Russell},
  {Lewis}, \& {Zdziarski}}]{Banerjee24}
{Banerjee}, S., {Dewangan}, G.~C., {Knigge}, C., {et~al.} 2024, \apj, 964, 189,
  \dodoi{10.3847/1538-4357/ad24ef}

\bibitem[{{Bhargava} {et~al.}(2021){Bhargava}, {Belloni}, {Bhattacharya},
  {Motta}, \& {Ponti.}}]{Bhargava21}
{Bhargava}, Y., {Belloni}, T., {Bhattacharya}, D., {Motta}, S., \& {Ponti.}, G.
  2021, \mnras, 508, 3104, \dodoi{10.1093/mnras/stab2848}

\bibitem[{{Bisnovatyi-Kogan} \& {Ruzmaikin}(1974)}]{BK74}
{Bisnovatyi-Kogan}, G.~S., \& {Ruzmaikin}, A.~A. 1974, \apss, 28, 45,
  \dodoi{10.1007/BF00642237}

\bibitem[{{Blandford} \& {K\"{o}nigl}(1979)}]{BK79}
{Blandford}, R.~D., \& {K\"{o}nigl}, A. 1979, \apj, 232, 34,
  \dodoi{10.1086/157262}

\bibitem[{{Blandford} \& {Znajek}(1977)}]{BZ77}
{Blandford}, R.~D., \& {Znajek}, R.~L. 1977, \mnras, 179, 433,
  \dodoi{10.1093/mnras/179.3.433}

\bibitem[{{Bright} {et~al.}(2020){Bright}, {Fender}, {Motta}, {Williams},
  {Moldon}, {Plotkin}, {Miller-Jones}, {Heywood}, {Tremou}, {Beswick},
  {Sivakoff}, {Corbel}, {Buckley}, {Homan}, {Gallo}, {Tetarenko}, {Russell},
  {Green}, {Titterington}, {Woudt}, {Armstrong}, {Groot}, {Horesh}, {van der
  Horst}, {K{\"o}rding}, {McBride}, {Rowlinson}, \& {Wijers}}]{Bright20}
{Bright}, J.~S., {Fender}, R.~P., {Motta}, S.~E., {et~al.} 2020, Nature
  Astronomy, 4, 697, \dodoi{10.1038/s41550-020-1023-5}

\bibitem[{{Carotenuto} {et~al.}(2024){Carotenuto}, {Fender}, {Tetarenko},
  {Zdziarski}, {Corbel}, {Shaik}, \& {Cooper}}]{Carotenuto24}
{Carotenuto}, F., {Fender}, R., {Tetarenko}, A.~J., {et~al.} 2024, in
  preparation

\bibitem[{{Carotenuto} {et~al.}(2022){Carotenuto}, {Tetarenko}, \&
  {Corbel}}]{Carotenuto22}
{Carotenuto}, F., {Tetarenko}, A.~J., \& {Corbel}, S. 2022, \mnras, 511, 4826,
  \dodoi{10.1093/mnras/stac329}

\bibitem[{{Carotenuto} {et~al.}(2021){Carotenuto}, {Corbel}, {Tremou},
  {Russell}, {Tzioumis}, {Fender}, {Woudt}, {Motta}, {Miller-Jones}, {Chauhan},
  {Tetarenko}, {Sivakoff}, {Heywood}, {Horesh}, {van der Horst}, {Koerding}, \&
  {Mooley}}]{Carotenuto21}
{Carotenuto}, F., {Corbel}, S., {Tremou}, E., {et~al.} 2021, \mnras, 504, 444,
  \dodoi{10.1093/mnras/stab864}

\bibitem[{{Casella} {et~al.}(2010){Casella}, {Maccarone}, {O'Brien}, {Fender},
  {Russell}, {van der Klis}, {Pe'Er}, {Maitra}, {Altamirano}, {Belloni},
  {Kanbach}, {Klein-Wolt}, {Mason}, {Soleri}, {Stefanescu}, {Wiersema}, \&
  {Wijnands}}]{Casella10}
{Casella}, P., {Maccarone}, T.~J., {O'Brien}, K., {et~al.} 2010, \mnras, 404,
  L21, \dodoi{10.1111/j.1745-3933.2010.00826.x}

\bibitem[{{Corbel} {et~al.}(2005){Corbel}, {Kaaret}, {Fender}, {Tzioumis},
  {Tomsick}, \& {Orosz}}]{Corbel05}
{Corbel}, S., {Kaaret}, P., {Fender}, R.~P., {et~al.} 2005, \apj, 632, 504,
  \dodoi{10.1086/432499}

\bibitem[{{Corral-Santana} {et~al.}(2016){Corral-Santana}, {Casares},
  {Mu{\~n}oz-Darias}, {Bauer}, {Mart{\'\i}nez-Pais}, \&
  {Russell}}]{Corral-Santana16}
{Corral-Santana}, J.~M., {Casares}, J., {Mu{\~n}oz-Darias}, T., {et~al.} 2016,
  \aap, 587, A61, \dodoi{10.1051/0004-6361/201527130}

\bibitem[{{Davis} \& {Tchekhovskoy}(2020)}]{Davis20}
{Davis}, S.~W., \& {Tchekhovskoy}, A. 2020, \araa, 58, 407,
  \dodoi{10.1146/annurev-astro-081817-051905}

\bibitem[{{Dhawan} {et~al.}(2000){Dhawan}, {Mirabel}, \&
  {Rodr{\'\i}guez}}]{Dhawan00}
{Dhawan}, V., {Mirabel}, I.~F., \& {Rodr{\'\i}guez}, L.~F. 2000, \apj, 543,
  373, \dodoi{10.1086/317088}

\bibitem[{{Done} {et~al.}(2007){Done}, {Gierli{\'n}ski}, \& {Kubota}}]{DGK07}
{Done}, C., {Gierli{\'n}ski}, M., \& {Kubota}, A. 2007, \aapr, 15, 1,
  \dodoi{10.1007/s00159-007-0006-1}

\bibitem[{{Espinasse} {et~al.}(2020){Espinasse}, {Corbel}, {Kaaret}, {Tremou},
  {Migliori}, {Plotkin}, {Bright}, {Tomsick}, {Tzioumis}, {Fender}, {Orosz},
  {Gallo}, {Homan}, {Jonker}, {Miller-Jones}, {Russell}, \&
  {Motta}}]{Espinasse20}
{Espinasse}, M., {Corbel}, S., {Kaaret}, P., {et~al.} 2020, \apjl, 895, L31,
  \dodoi{10.3847/2041-8213/ab88b6}

\bibitem[{{Fabian} {et~al.}(2020){Fabian}, {Buisson}, {Kosec}, {Reynolds},
  {Wilkins}, {Tomsick}, {Walton}, {Gandhi}, {Altamirano}, {Arzoumanian},
  {Cackett}, {Dyda}, {Garcia}, {Gendreau}, {Grefenstette}, {Homan}, {Kara},
  {Ludlam}, {Miller}, \& {Steiner}}]{Fabian20}
{Fabian}, A.~C., {Buisson}, D.~J., {Kosec}, P., {et~al.} 2020, \mnras, 493,
  5389, \dodoi{10.1093/mnras/staa564}

\bibitem[{{Fanaroff} \& {Riley}(1974)}]{FR74}
{Fanaroff}, B.~L., \& {Riley}, J.~M. 1974, \mnras, 167, 31P,
  \dodoi{10.1093/mnras/167.1.31P}

\bibitem[{{Fender}(2001)}]{Fender01b}
{Fender}, R.~P. 2001, \mnras, 322, 31, \dodoi{10.1046/j.1365-8711.2001.04080.x}

\bibitem[{{Fender} {et~al.}(2004){Fender}, {Belloni}, \& {Gallo}}]{FBG04}
{Fender}, R.~P., {Belloni}, T.~M., \& {Gallo}, E. 2004, \mnras, 355, 1105,
  \dodoi{10.1111/j.1365-2966.2004.08384.x}

\bibitem[{{Fender} {et~al.}(1999){Fender}, {Garrington}, {McKay}, {Muxlow},
  {Pooley}, {Spencer}, {Stirling}, \& {Waltman}}]{Fender99}
{Fender}, R.~P., {Garrington}, S.~T., {McKay}, D.~J., {et~al.} 1999, \mnras,
  304, 865, \dodoi{10.1046/j.1365-8711.1999.02364.x}

\bibitem[{{Fragile} {et~al.}(2023){Fragile}, {Chatterjee}, {Ingram}, \&
  {Middleton}}]{Fragile23}
{Fragile}, P.~C., {Chatterjee}, K., {Ingram}, A., \& {Middleton}, M. 2023,
  \mnras, 525, L82, \dodoi{10.1093/mnrasl/slad099}

\bibitem[{{Gallo} {et~al.}(2005){Gallo}, {Fender}, {Kaiser}, {Russell},
  {Morganti}, {Oosterloo}, \& {Heinz}}]{Gallo05}
{Gallo}, E., {Fender}, R., {Kaiser}, C., {et~al.} 2005, \nat, 436, 819,
  \dodoi{10.1038/nature03879}

\bibitem[{{Guan} {et~al.}(2021){Guan}, {Tao}, {Qu}, {Zhang}, {Zhang}, {Zhang},
  {Ma}, {Ge}, {Song}, {Lu}, {Li}, {Xu}, {Chen}, {Cao}, {Liu}, {Zhang}, {Wang},
  {Chen}, {Bu}, {Cai}, {Chang}, {Chen}, {Chen}, {Chen}, {Cui}, {Du}, {Gao},
  {Gao}, {Gu}, {Guo}, {Han}, {Huang}, {Huo}, {Jia}, {Jiang}, {Jin}, {Kong},
  {Li}, {Li}, {Li}, {Li}, {Li}, {Li}, {Li}, {Li}, {Liang}, {Liao}, {Liu},
  {Liu}, {Liu}, {Liu}, {Lu}, {Luo}, {Luo}, {Ma}, {Meng}, {Nang}, {Nie}, {Ou},
  {Ren}, {Sai}, {Song}, {Sun}, {Tan}, {Wang}, {Wang}, {Wang}, {Wang}, {Wang},
  {Wen}, {Wu}, {Wu}, {Wu}, {Xiao}, {Xiao}, {Xiong}, {Yang}, {Yang}, {Yang},
  {Yang}, {Yi}, {Yin}, {You}, {Zhang}, {Zhang}, {Zhang}, {Zhang}, {Zhang},
  {Zhang}, {Zhang}, {Zhao}, {Zhao}, {Zheng}, {Zheng}, \& {Zhou}}]{Guan21}
{Guan}, J., {Tao}, L., {Qu}, J.~L., {et~al.} 2021, \mnras, 504, 2168,
  \dodoi{10.1093/mnras/stab945}

\bibitem[{{Heinz}(2002)}]{Heinz02}
{Heinz}, S. 2002, \aap, 388, L40, \dodoi{10.1051/0004-6361:20020402}

\bibitem[{{Kaiser}(2006)}]{Kaiser06}
{Kaiser}, C.~R. 2006, \mnras, 367, 1083,
  \dodoi{10.1111/j.1365-2966.2006.10030.x}

\bibitem[{{Kawamuro} {et~al.}(2018){Kawamuro}, {Negoro}, {Yoneyama}, {Ueno},
  {Tomida}, {Ishikawa}, {Sugawara}, {Isobe}, {Shimomukai}, {Mihara},
  {Sugizaki}, {Nakahira}, {Iwakiri}, {Yatabe}, {Takao}, {Matsuoka}, {Kawai},
  {Sugita}, {Yoshii}, {Tachibana}, {Harita}, {Morita}, {Yoshida}, {Sakamoto},
  {Serino}, {Kawakubo}, {Kitaoka}, {Hashimoto}, {Tsunemi}, {Nakajima},
  {Kawase}, {Sakamaki}, {Maruyama}, {Ueda}, {Hori}, {Tanimoto}, {Oda},
  {Morita}, {Yamada}, {Tsuboi}, {Nakamura}, {Sasaki}, {Kawai}, {Sato},
  {Yamauchi}, {Hanyu}, {Hidaka}, {Yamaoka}, \& {Shidatsu}}]{Kawamuro18}
{Kawamuro}, T., {Negoro}, H., {Yoneyama}, T., {et~al.} 2018, Astron.\ Telegram,
  11399, 1

\bibitem[{{Koljonen} {et~al.}(2010){Koljonen}, {Hannikainen}, {McCollough},
  {Pooley}, \& {Trushkin}}]{Koljonen10}
{Koljonen}, K.~I.~I., {Hannikainen}, D.~C., {McCollough}, M.~L., {Pooley},
  G.~G., \& {Trushkin}, S.~A. 2010, \mnras, 406, 307,
  \dodoi{10.1111/j.1365-2966.2010.16722.x}

\bibitem[{{Luque-Escamilla} {et~al.}(2015){Luque-Escamilla}, {Mart{\'\i}}, \&
  {Mart{\'\i}nez-Aroza}}]{Luque15}
{Luque-Escamilla}, P.~L., {Mart{\'\i}}, J., \& {Mart{\'\i}nez-Aroza}, J. 2015,
  \aap, 584, A122, \dodoi{10.1051/0004-6361/201527238}

\bibitem[{{Mart{\'\i}} {et~al.}(2017){Mart{\'\i}}, {Luque-Escamilla},
  {Bosch-Ramon}, \& {Paredes}}]{Marti17}
{Mart{\'\i}}, J., {Luque-Escamilla}, P.~L., {Bosch-Ramon}, V., \& {Paredes},
  J.~M. 2017, Nature Communications, 8, 1757,
  \dodoi{10.1038/s41467-017-01976-5}

\bibitem[{{Mart{\'\i}} {et~al.}(2002){Mart{\'\i}}, {Mirabel}, {Rodr{\'\i}guez},
  \& {Smith}}]{Marti02}
{Mart{\'\i}}, J., {Mirabel}, I.~F., {Rodr{\'\i}guez}, L.~F., \& {Smith}, I.~A.
  2002, \aap, 386, 571, \dodoi{10.1051/0004-6361:20020273}

\bibitem[{{McKinney} {et~al.}(2012){McKinney}, {Tchekhovskoy}, \&
  {Blandford}}]{McKinney12}
{McKinney}, J.~C., {Tchekhovskoy}, A., \& {Blandford}, R.~D. 2012, \mnras, 423,
  3083, \dodoi{10.1111/j.1365-2966.2012.21074.x}

\bibitem[{{Miko{\l}ajewska} {et~al.}(2022){Miko{\l}ajewska}, {Zdziarski},
  {Zi{\'o}{\l}kowski}, {Torres}, \& {Casares}}]{Mikolajewska22}
{Miko{\l}ajewska}, J., {Zdziarski}, A.~A., {Zi{\'o}{\l}kowski}, J., {Torres},
  M. A.~P., \& {Casares}, J. 2022, \apj, 930, 9,
  \dodoi{10.3847/1538-4357/ac6099}

\bibitem[{{Mirabel} \& {Rodr{\'{\i}}guez}(1994)}]{MR94}
{Mirabel}, I.~F., \& {Rodr{\'{\i}}guez}, L.~F. 1994, \nat, 371, 46,
  \dodoi{10.1038/371046a0}

\bibitem[{{Mirabel} {et~al.}(1992){Mirabel}, {Rodriguez}, {Cordier}, {Paul}, \&
  {Lebrun}}]{Mirabel92}
{Mirabel}, I.~F., {Rodriguez}, L.~F., {Cordier}, B., {Paul}, J., \& {Lebrun},
  F. 1992, \nat, 358, 215, \dodoi{10.1038/358215a0}

\bibitem[{{Narayan} {et~al.}(2003){Narayan}, {Igumenshchev}, \&
  {Abramowicz}}]{Narayan03}
{Narayan}, R., {Igumenshchev}, I.~V., \& {Abramowicz}, M.~A. 2003, \pasj, 55,
  L69, \dodoi{10.1093/pasj/55.6.L69}

\bibitem[{{Narayan} {et~al.}(2012){Narayan}, {S{\k{a}}dowski}, {Penna}, \&
  {Kulkarni}}]{Narayan12b}
{Narayan}, R., {S{\k{a}}dowski}, A., {Penna}, R.~F., \& {Kulkarni}, A.~K. 2012,
  \mnras, 426, 3241, \dodoi{10.1111/j.1365-2966.2012.22002.x}

\bibitem[{{Prabu} {et~al.}(2023){Prabu}, {Miller-Jones}, {Bahramian}, {Wood},
  {Tingay}, {Atri}, {Plotkin}, \& {Strader}}]{Prabu23}
{Prabu}, S., {Miller-Jones}, J.~C.~A., {Bahramian}, A., {et~al.} 2023, \mnras,
  525, 4426, \dodoi{10.1093/mnras/stad2570}

\bibitem[{{Reid} {et~al.}(2014){Reid}, {McClintock}, {Steiner}, {Steeghs},
  {Remillard}, {Dhawan}, \& {Narayan}}]{Reid14}
{Reid}, M.~J., {McClintock}, J.~E., {Steiner}, J.~F., {et~al.} 2014, \apj, 796,
  2, \dodoi{10.1088/0004-637X/796/1/2}

\bibitem[{{Rodi} {et~al.}(2021){Rodi}, {Tramacere}, {Onori}, {Bruni},
  {S{\`a}nchez-Fern{\`a}ndez}, {Fiocchi}, {Natalucci}, \& {Ubertini}}]{Rodi21}
{Rodi}, J., {Tramacere}, A., {Onori}, F., {et~al.} 2021, \apj, 910, 21,
  \dodoi{10.3847/1538-4357/abdfd0}

\bibitem[{{Rodriguez} {et~al.}(1995){Rodriguez}, {Gerard}, {Mirabel}, {Gomez},
  \& {Velazquez}}]{Rodriguez95}
{Rodriguez}, L.~F., {Gerard}, E., {Mirabel}, I.~F., {Gomez}, Y., \&
  {Velazquez}, A. 1995, \apjs, 101, 173, \dodoi{10.1086/192236}

\bibitem[{{Rodriguez} {et~al.}(1992){Rodriguez}, {Mirabel}, \&
  {Marti}}]{Rodriguez92}
{Rodriguez}, L.~F., {Mirabel}, I.~F., \& {Marti}, J. 1992, \apjl, 401, L15,
  \dodoi{10.1086/186659}

\bibitem[{{Rushton} {et~al.}(2017){Rushton}, {Miller-Jones}, {Curran},
  {Sivakoff}, {Rupen}, {Paragi}, {Spencer}, {Yang}, {Altamirano}, {Belloni},
  {Fender}, {Krimm}, {Maitra}, {Migliari}, {Russell}, {Russell}, {Soria}, \&
  {Tudose}}]{Rushton17}
{Rushton}, A.~P., {Miller-Jones}, J.~C.~A., {Curran}, P.~A., {et~al.} 2017,
  \mnras, 468, 2788, \dodoi{10.1093/mnras/stx526}

\bibitem[{{Russell} {et~al.}(2015){Russell}, {Miller-Jones}, {Curran}, {Soria},
  {Altamirano}, {Corbel}, {Coriat}, {Moin}, {Russell}, {Sivakoff},
  {Slaven-Blair}, {Belloni}, {Fender}, {Heinz}, {Jonker}, {Krimm},
  {K{\"o}rding}, {Maitra}, {Markoff}, {Middleton}, {Migliari}, {Remillard},
  {Rupen}, {Sarazin}, {Tetarenko}, {Torres}, {Tudose}, \&
  {Tzioumis}}]{Russell15}
{Russell}, T.~D., {Miller-Jones}, J.~C.~A., {Curran}, P.~A., {et~al.} 2015,
  \mnras, 450, 1745, \dodoi{10.1093/mnras/stv723}

\bibitem[{{Sell} {et~al.}(2015){Sell}, {Heinz}, {Richards}, {Maccarone},
  {Russell}, {Gallo}, {Fender}, {Markoff}, \& {Nowak}}]{Sell15}
{Sell}, P.~H., {Heinz}, S., {Richards}, E., {et~al.} 2015, \mnras, 446, 3579,
  \dodoi{10.1093/mnras/stu2320}

\bibitem[{{Shakura} \& {Sunyaev}(1973)}]{SS73}
{Shakura}, N.~I., \& {Sunyaev}, R.~A. 1973, \aap, 24, 337

\bibitem[{{Shidatsu} {et~al.}(2019){Shidatsu}, {Nakahira}, {Murata}, {Adachi},
  {Kawai}, {Ueda}, \& {Negoro}}]{Shidatsu19}
{Shidatsu}, M., {Nakahira}, S., {Murata}, K.~L., {et~al.} 2019, \apj, 874, 183,
  \dodoi{10.3847/1538-4357/ab09ff}

\bibitem[{{Sikora} {et~al.}(2020){Sikora}, {Nalewajko}, \&
  {Madejski}}]{Sikora20}
{Sikora}, M., {Nalewajko}, K., \& {Madejski}, G.~M. 2020, \mnras, 499, 3749,
  \dodoi{10.1093/mnras/staa3128}

\bibitem[{{Sikora} \& {Zdziarski}(2023)}]{SZ23}
{Sikora}, M., \& {Zdziarski}, A.~A. 2023, \apjl, 954, L30,
  \dodoi{10.3847/2041-8213/acf1a0}

\bibitem[{{Steiner} \& {McClintock}(2012)}]{Steiner12}
{Steiner}, J.~F., \& {McClintock}, J.~E. 2012, \apj, 745, 136,
  \dodoi{10.1088/0004-637X/745/2/136}

\bibitem[{{Steiner} {et~al.}(2012){Steiner}, {McClintock}, \&
  {Reid}}]{Steiner12b}
{Steiner}, J.~F., {McClintock}, J.~E., \& {Reid}, M.~J. 2012, \apjl, 745, L7,
  \dodoi{10.1088/2041-8205/745/1/L7}

\bibitem[{{Stirling} {et~al.}(2001){Stirling}, {Spencer}, {de la Force},
  {Garrett}, {Fender}, \& {Ogley}}]{Stirling01}
{Stirling}, A.~M., {Spencer}, R.~E., {de la Force}, C.~J., {et~al.} 2001,
  \mnras, 327, 1273, \dodoi{10.1046/j.1365-8711.2001.04821.x}

\bibitem[{{Tchekhovskoy}(2015)}]{Tchekhovskoy15}
{Tchekhovskoy}, A. 2015, {Launching of Active Galactic Nuclei Jets}, Vol. 414
  (Springer), 45, \dodoi{10.1007/978-3-319-10356-3\_3}

\bibitem[{{Tchekhovskoy} {et~al.}(2011){Tchekhovskoy}, {Narayan}, \&
  {McKinney}}]{Tchekhovskoy11}
{Tchekhovskoy}, A., {Narayan}, R., \& {McKinney}, J.~C. 2011, \mnras, 418, L79,
  \dodoi{10.1111/j.1745-3933.2011.01147.x}

\bibitem[{{Tetarenko} {et~al.}(2019){Tetarenko}, {Casella}, {Miller-Jones},
  {Sivakoff}, {Tetarenko}, {Maccarone}, {Gandhi}, \&
  {Eikenberry}}]{Tetarenko19}
{Tetarenko}, A.~J., {Casella}, P., {Miller-Jones}, J.~C.~A., {et~al.} 2019,
  \mnras, 484, 2987, \dodoi{10.1093/mnras/stz165}

\bibitem[{{Tetarenko} {et~al.}(2021){Tetarenko}, {Casella}, {Miller-Jones},
  {Sivakoff}, {Paice}, {Vincentelli}, {Maccarone}, {Gandhi}, {Dhillon}, \&
  {Marsh}}]{Tetarenko21}
---. 2021, \mnras, 504, 3862, \dodoi{10.1093/mnras/stab820}

\bibitem[{{Thomas} {et~al.}(2022){Thomas}, {Charles}, {Buckley}, {Kotze},
  {Lasota}, {Potter}, {Steiner}, \& {Paice}}]{Thomas22}
{Thomas}, J.~K., {Charles}, P.~A., {Buckley}, D. A.~H., {et~al.} 2022, \mnras,
  509, 1062, \dodoi{10.1093/mnras/stab3033}

\bibitem[{{Torres} {et~al.}(2020){Torres}, {Casares}, {Jim{\'e}nez-Ibarra},
  {{\'A}lvarez-Hern{\'a}ndez}, {Mu{\~n}oz-Darias}, {Armas Padilla}, {Jonker},
  \& {Heida}}]{Torres20}
{Torres}, M.~A.~P., {Casares}, J., {Jim{\'e}nez-Ibarra}, F., {et~al.} 2020,
  \apjl, 893, L37, \dodoi{10.3847/2041-8213/ab863a}

\bibitem[{{Wang} {et~al.}(2003){Wang}, {Dai}, \& {Lu}}]{Wang03}
{Wang}, X.~Y., {Dai}, Z.~G., \& {Lu}, T. 2003, \apj, 592, 347,
  \dodoi{10.1086/375638}

\bibitem[{{Wood} {et~al.}(2021){Wood}, {Miller-Jones}, {Homan}, {Bright},
  {Motta}, {Fender}, {Markoff}, {Belloni}, {K{\"o}rding}, {Maitra}, {Migliari},
  {Russell}, {Russell}, {Sarazin}, {Soria}, {Tetarenko}, \& {Tudose}}]{Wood21}
{Wood}, C.~M., {Miller-Jones}, J.~C.~A., {Homan}, J., {et~al.} 2021, \mnras,
  505, 3393, \dodoi{10.1093/mnras/stab1479}

\bibitem[{{Yuan} \& {Narayan}(2014)}]{YN14}
{Yuan}, F., \& {Narayan}, R. 2014, \araa, 52, 529,
  \dodoi{10.1146/annurev-astro-082812-141003}

\bibitem[{{Yuan} {et~al.}(2018){Yuan}, {Wang}, {Worrall}, {Zhang}, \&
  {Mao}}]{Yuan18}
{Yuan}, Z., {Wang}, J., {Worrall}, D.~M., {Zhang}, B.-B., \& {Mao}, J. 2018,
  \apjs, 239, 33, \dodoi{10.3847/1538-4365/aaed3b}

\bibitem[{{Zdziarski}(2014)}]{Zdziarski14d}
{Zdziarski}, A.~A. 2014, \mnras, 444, 1113, \dodoi{10.1093/mnras/stu1525}

\bibitem[{{Zdziarski} \& {Egron}(2022)}]{Z_Egron22}
{Zdziarski}, A.~A., \& {Egron}, E. 2022, \apjl, 935, L4,
  \dodoi{10.3847/2041-8213/ac81bf}

\bibitem[{{Zdziarski} {et~al.}(2022{\natexlab{a}}){Zdziarski}, {Phuravhathu},
  {Sikora}, {B{\"o}ttcher}, \& {Chibueze}}]{Zdziarski22c}
{Zdziarski}, A.~A., {Phuravhathu}, D.~G., {Sikora}, M., {B{\"o}ttcher}, M., \&
  {Chibueze}, J.~O. 2022{\natexlab{a}}, \apjl, 928, L9,
  \dodoi{10.3847/2041-8213/ac5b70}

\bibitem[{{Zdziarski} {et~al.}(2023){Zdziarski}, {Sikora}, {Szanecki}, \&
  {B{\"o}ttcher}}]{Zdziarski23a}
{Zdziarski}, A.~A., {Sikora}, M., {Szanecki}, M., \& {B{\"o}ttcher}, M. 2023,
  \apjl, 947, L32, \dodoi{10.3847/2041-8213/accb5a}

\bibitem[{{Zdziarski} {et~al.}(2022{\natexlab{b}}){Zdziarski}, {Tetarenko}, \&
  {Sikora}}]{Zdziarski22a}
{Zdziarski}, A.~A., {Tetarenko}, A.~J., \& {Sikora}, M. 2022{\natexlab{b}},
  \apj, 925, 189, \dodoi{10.3847/1538-4357/ac38a9}

\bibitem[{{Zhao} {et~al.}(2021){Zhao}, {Gou}, {Dong}, {Tuo}, {Liao}, {Li},
  {Jia}, {Feng}, \& {Steiner}}]{Zhao21}
{Zhao}, X., {Gou}, L., {Dong}, Y., {et~al.} 2021, \apj, 916, 108,
  \dodoi{10.3847/1538-4357/ac07a9}

\end{thebibliography}

\end{document}